\documentclass[10pt, a4paper,twocolumn]{article}

\usepackage{graphicx}
\usepackage{hyperref}
\usepackage{color}
\usepackage[table]{xcolor}
\usepackage{multirow}
\usepackage{booktabs} 
\usepackage[english]{babel} 
\usepackage{enumerate}
\usepackage{amssymb,amsmath,amsthm, amsfonts}

\RequirePackage[left=2cm,%
                right=2cm,%
                top=2.25cm,%
                bottom=2.25cm,%
                headheight=12pt,%
                letterpaper]{geometry}%

\theoremstyle{definition}
\newcommand\Tstrut{\rule{0pt}{2.6ex}}  

\usepackage[normalem]{ulem} 
\hypersetup{breaklinks=true}
\hypersetup{
	colorlinks,
	linkcolor={blue!50!black},
	citecolor={blue!50!black},
	urlcolor={blue!50!black}
}
\usepackage{authblk}
\usepackage[numbers,sort&compress]{natbib}

\title{Exploiting ergodicity of the logistic map using deep-zoom to improve security of chaos-based cryptosystems}
\author[1]{Jeaneth Machicao}
\author[2]{Marcela Alves}
\author[3]{Murilo S. Baptista}
\author[1,2]{Odemir M. Bruno}
\affil[1]{\small{S\~{a}o Carlos Institute of Physics, University of S\~{a}o Paulo (USP), PO Box 369, 13560-970, S\~{a}o Carlos, SP, Brazil. \protect\\Scientific Computing Group - http://scg.ifsc.usp.br}}
\affil[2]{\small{Institute of Mathematics and Computer Science, University of S\~{a}o Paulo (USP), USP, Avenida Trabalhador s\~ao-carlense, 400, 13566-590, S\~ao Carlos, SP, Brazil.}}
\affil[3]{\small{University of Aberdeen, Institute of Complex Systems and Mathematical Biology, AB24 3UE, Aberdeen, UK}}
\date{\vspace{-5ex}}

\begin{document}
\maketitle

\begin{abstract}
This paper explores the deep-zoom properties of the chaotic $k$-logistic map, in order to propose an improved chaos-based cryptosystem. This map was shown to enhance the random features of the Logistic map, while at the same time reducing the predictability about its orbits. We incorporate its strengths to security into a previously published cryptosystem to provide an optimal pseudo-random number generator (PRNG) as its core operation. The result is a reliable method that does not have the weaknesses previously reported about the original cryptosystem. 
\end{abstract}

\section{Introduction}

The emergence of the application of chaos theory in cryptographic algorithms has caught the attention of many researchers since the 1980s \cite{Baptista,Kocarev2001,kocarev2004public,Patidar,szczepanski2005cryptographically,Alvarez2006,masuda2006chaotic,Anderson2010,Pisarchik2010,MachicaoLifeLike,Vidal2014,Lin2016,MachicaoHash}. Chaos has two faces. It is a deterministic system, but if correctly observed can behave as a random one. This stochastic feature of the chaotic systems have makes it interesting to its use in cryptography. Because its random nature can emerge even from relatively simple equations, chaos is suitable to create efficient pseudo-random number generators (PRNG). Chaotic systems are sensitivity to the initial conditions, and thus predictability is limited to a short time window. They are mixing, and thus information about future states cannot reveal past states. These two properties are intrinsically connected to the random nature of chaos. They are also transitive and have an infinitude of unstable periodic orbits, and thus present infinitely recurrent patterns. This later property is intrinsically linked to the deterministic nature of chaos. Chaos-based cryptosystems explore the random properties of chaos to secure information. However, its deterministic nature also allows for a unique cryptoanalysis that usually explores the short-term predictability and periodicity of chaos \cite{AlvarezCriptonalise, Persohn2012}.

One way to turn chaotic systems into magnificent semi-random systems would be to reduce its short-term predictability and periodicity, but without having to rely on long-term iterates of it. This was recently achieved by considering the deep-zoom approach to chaotic maps \cite{machicao2017}, which consists to improve the pseudo-randomness quality of another chaotic orbit by removing $k$-digits to the right of the decimal point from each point of the original orbit. In this manner, the emergence of random-style patterns occurs as finer computations are performed. Peeking into less significant digits can reveal a never before explored pseudo-randomness properties of chaotic systems. The patterns and the predictability of the chaotic maps can change this behavior according to the scale that the orbits are observed. 

We consider a chaos-based method published in the late 1990s \cite{Baptista}, which is an interesting object of study due to its simplicity of implementation, dissemination in the academic community, the number of published works presenting mainly the failures mentioned above, and that has a proven security-failure behavior \cite{AlvarezCriptonalise}, thus an optimal candidate to the introduction a novel approach for security. The main flaws presented in \citeauthor{Baptista}'s work are directly related to the statistics of the recurrent orbits and their predictability in the logistic map. In this work, we maintain the essence of Baptista's original approach, which explores the statistical feature of the orbits recurrence to secure information. However, we use the $k$-logistic map, instead of the logistic map used in the original approach.

The cryptology field is in a continuous battle to yield strong encryption methods (cryptography) and to find weaknesses aiming to break these ciphers (cryptoanalysis). This establishes a recurrent cycle in which novel breakthrough cryptographic methods are being released, and eventually, being scrutinized by cryptanalysts in order to report possible flaws. We must emphasize that, we do not aim to construct a cryptographically safe algorithm but instead to focus on the theoretical study of the proposed improvement version. This paper shows that the \citeauthor{Baptista}-like cryptosystem can be embedded with the pseudo-randomness sources of the $k$-logistic map, as this PRNG has shown to pass several tests that the original system has failed. Thus, we propose a manner to improve the main flaws for which an already broken method that uses a well-known chaotic system has been deprecated, opening up a challenge for the scientific community to cryptanalyze it. 

In Sec. \ref{sec:Baptista}, we introduce the original Baptista's method, describing its parameters and procedures. In Sec. 
\ref{sec3}, we introduce the logistic map and explain the deep-zoom property of chaos, leading to the definition of the $k$-logistic map. Then, in Sec. \ref{sec4}, we introduce our proposed cryptosystem, followed in Sec. \ref{sec:analysis} by its cryptanalysis. Discussions and conclusions are presented in Sec. \ref{sec6}. 

\section{Baptista's chaos-based cryptographic system}
\label{sec:Baptista} 

\citeauthor{Baptista}'s algorithm is one of the first digital chaos-based cryptographic system introduced into the state-of-art \cite{Pisarchik2010}. This algorithm takes advantage of the ergodic property of simple and low-dimensional logistic equation in order to create a ciphertext.

The logistic map is a quadratic function of the form $f:[0,1]\rightarrow [0,1]$ given by Eq.~\eqref{eq:logisticmap}.

\begin{equation}
\label{eq:logisticmap}
x_{n+1}= f(x_{n}) = \mu x_n(1 - x_n) \,,
\end{equation}

\noindent where $x_n \in [0,1]$ and $\mu \in [0, 4]$, showing chaotic behavior depending on the value of the parameter $\mu$. This map is used as a paradigm for nonlinear dynamical systems due to its simplicity, speed \cite{Pisarchik2010}, and well-known behavior and because it is extensively applied in the areas of physics, biology, mathematics, chemistry and others \cite{elhadj2008robustness}.

\begin{figure}[!ht]
	\begin{center}
		\includegraphics[scale=0.8]{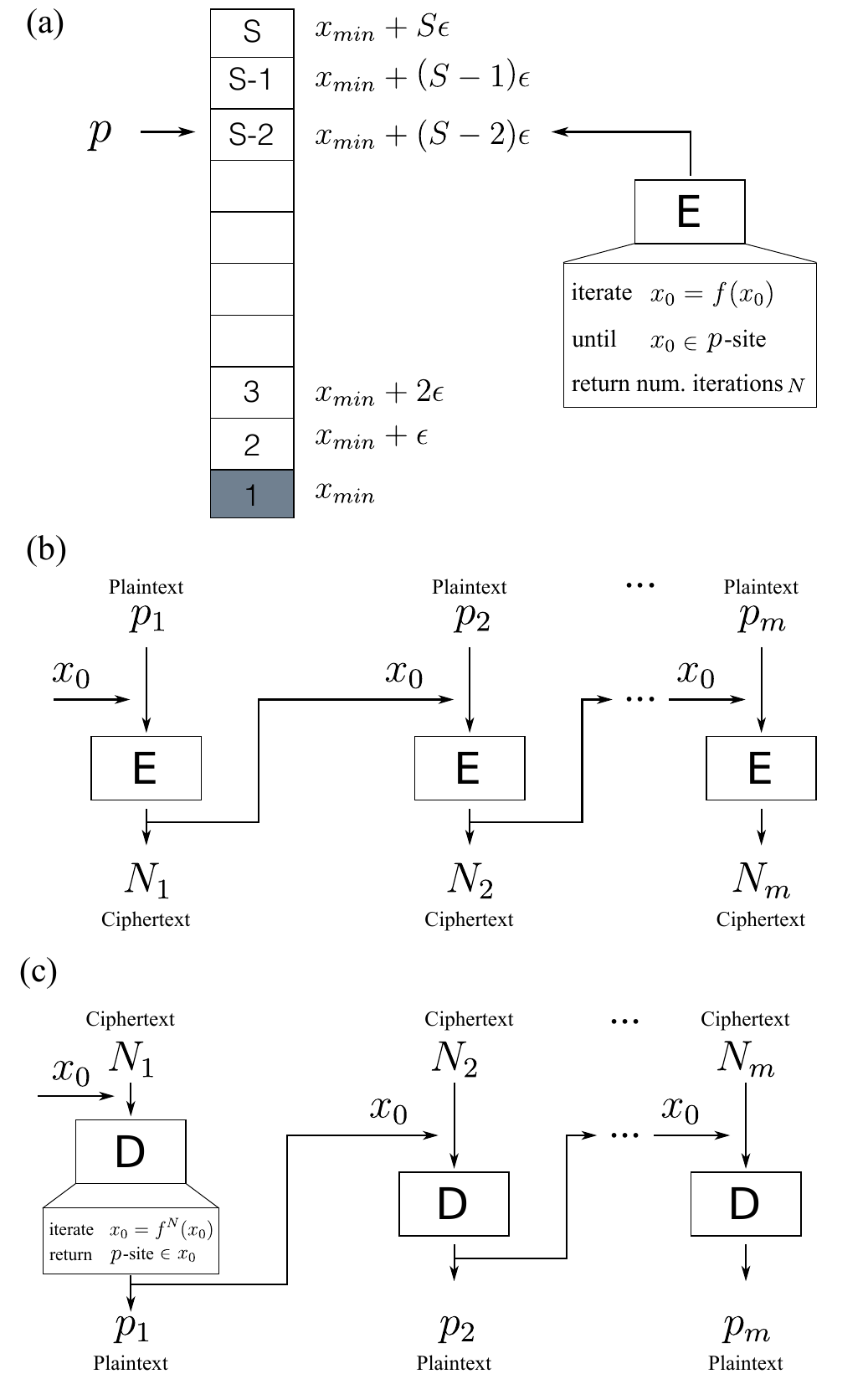}
		\caption{Illustration of the encryption/decryption process of the Baptista chaos-based cryptographic algorithm \cite{Baptista}. a) The phase space of the logistic map is divided in $S$ sites, each one with size $E = (x_{\text{max}} - x_{\text{min}})/S$, where $x_{\text{min}}$ and $x_{\text{max}}$ are the limit ranges of the phase space being considered. For each site an alphabet unit $p$ is associated. b) The encryption process describe the number of iterations $N$ the logistic map (Eq.~\eqref{eq:logisticmap}) must iterate until $x_0$ achieves the $p$-site range, then $x_0$ is passed to encrypt the next plaintext unit. c) The decryption process implies to iterate the logistic map over $N$ iterations and return the corresponding $p$-site.}
		\label{fig:baptistaAlgorithm}
	\end{center}
\end{figure}

This algorithm associates each letter of the alphabet with a $\varepsilon$-interval of the phase space of the logistic map. Fig.~\ref{fig:baptistaAlgorithm}a is a representation of how the phase space is divided in $S$ sites, $x_{\text{min}}$ and $x_{\text{max}}$ are the minimum and maximum boundary of the phase space, and the size of the $\varepsilon$-intervals depends on the parameter $S$, representing the number of different letters in an alphabet one wants to encode a message. According to the detailed procedures \cite{Baptista}, it is possible to encrypt a message so that each letter of the message is encrypted as the number of iterations applied in Eq.~\eqref{eq:logisticmap}. Thereby each letter is represented by an integer number of iterations performed by the logistic equation, so the trajectory goes from an initial condition $x_0$ and reaches a $\varepsilon$-interval associated with that letter. The sender defines these associations in order to create a key.

The algorithm depends on the association of $\varepsilon$-intervals with the letters of the alphabet, the definition of the first initial condition, $x_0$ and a choice of a control parameter $\mu$, causing a chaotic behavior in the logistic map. The ergodic property causes the event of a $\varepsilon$-interval being reached by an infinite number of trajectories. For this reason, it is considered the $N$-iteration sized trajectory. It is also important to point out that a transient time $N_0$ will represent the minimum length that any encoding trajectory should have.

The original algorithm was implemented using trajectories with less than 65 532 iterations \cite{Baptista}. Even with this size restriction, the encrypted text can still be represented by different combinations, therefore the sender has to choose one in a stochastic fashion. In this way, the sender also defines a coefficient $\eta$, which represents a probability with which the receiver/transmitter picks a particular return as the encoding/decoding message.

The procedure given in the original paper \cite{Baptista} is replicated in Table~\ref{tab:batistatransmitter}. It exemplifies the encryption of the word ``hi''. The parameters are $\mu=3.8$, $S=256$, $\varepsilon = 0.00234375$, and the limits of the attractor are $x_{\text{min}}=0.2$ and $x_{\text{max}}=0.8$. The transient time is $N_0 = 250$ and the coefficient is $\eta=0$. The alphabet with $S=256$ letters, accordingly, it is defined 256 $\varepsilon$-intervals. It considers trajectories that have lengths smaller or equal to 65 532, but that are larger than $N_0$, which is sufficient to allow that a typical point of a $\epsilon$-interval of the order of 1/256 reaches out any other equally-sized $\epsilon$-interval several times. This can be guaranteed knowing that for any two arbitrary $\epsilon$ intervals of the logistic map, the shortest time for a trajectory point to leave an interval and arrive at another one can be estimated by the Lyapunov time, and given by $-1/\lambda \log{(\epsilon)}$. The coefficient $\eta$ also represents the probability for the transmitter to adapt the length of a trajectory visiting the desired interval as the ciphertext. 

The encryption procedure is a one-by-one transformation as shown in Fig.~\ref{fig:baptistaAlgorithm}b, i.e., the initial condition changes for each unit (letter) of the plaintext $P_1=$``h'' and $P_2=$``i'', which are the ASCII letters associated to the split units 104 and 105, respectively. In this way, if $C_1$ is the first unit of the ciphertext, and $C_2$ would be the second unit of the ciphertext, it was generated by the $x_0$ defined by the transmitter. In order to generate the third unit, $C_2$ is used as a new initial condition. Therefore, the message sent is $C_1=1713$ and $C_2=364$.

As seen in Fig.~\ref{fig:baptistaAlgorithm}c, the receiver may decrypt the message since the ciphertext indicates how many times Eq.~\eqref{eq:logisticmap} has to be iterated in order to find the corresponding letter. Following the previous examples shown in Table~\ref{tab:batistareceiver}, the logistic equation must be iterated 1713 times. The result is the value of $x$ associated with the interval 104, the interval that represents the letter ``h''. This new value of $x$ is considered as $x_0$ for the next step, the receiver iterates the logistic map equation 364 times, reaching the interval 105, used to represent the letter ``i''.

\begin{table}[!ht]
	\caption{Encryption procedure of the ergodicity chaos-based algorithm. From left to right the plaintext ``hi'' associated to their corresponding site number based on an initial condition $x_0$, which is iterated over $x$ time steps until returns a ciphertext $C_n$.}
	\label{tab:batistatransmitter} 
	\small
		\begin{tabular}{lllll}
			\toprule
			Plain & $x_0$ & $x_n$ & Site & $C_n$\\
			\hline
			\addlinespace[1ex]
			``h'' & 0.23232300000000 & 0.44160905447136& 104 & 1713 \\
			``i'' & 0.44160905447136 & 0.44486572362642& 105 & 364 \\
			\bottomrule
		\end{tabular}  
\end{table}

\begin{table}[!ht]
	\caption{Decryption procedure of the ergodicity chaos-based algorithm.}
	\label{tab:batistareceiver}
	\small
		\begin{tabular}{lllll}
		\toprule
			\textbf{$C_n$} & $x_0$ & $x_n$ & Site & Plain\\
			\hline
			\addlinespace[1ex]
			1713 & 0.23232300000000 & 0.44160905447136& 104 & ``h'' \\
			364 & 0.44160905447136 & 0.44486572362642&105 & ``i'' \\
\bottomrule			
		\end{tabular} 
\end{table}

\section{The logistic map as a source of randomness}\label{sec3}

\subsection{Logistic map}
\begin{figure*}[!ht]
	\centering 
	\includegraphics[scale=0.78]{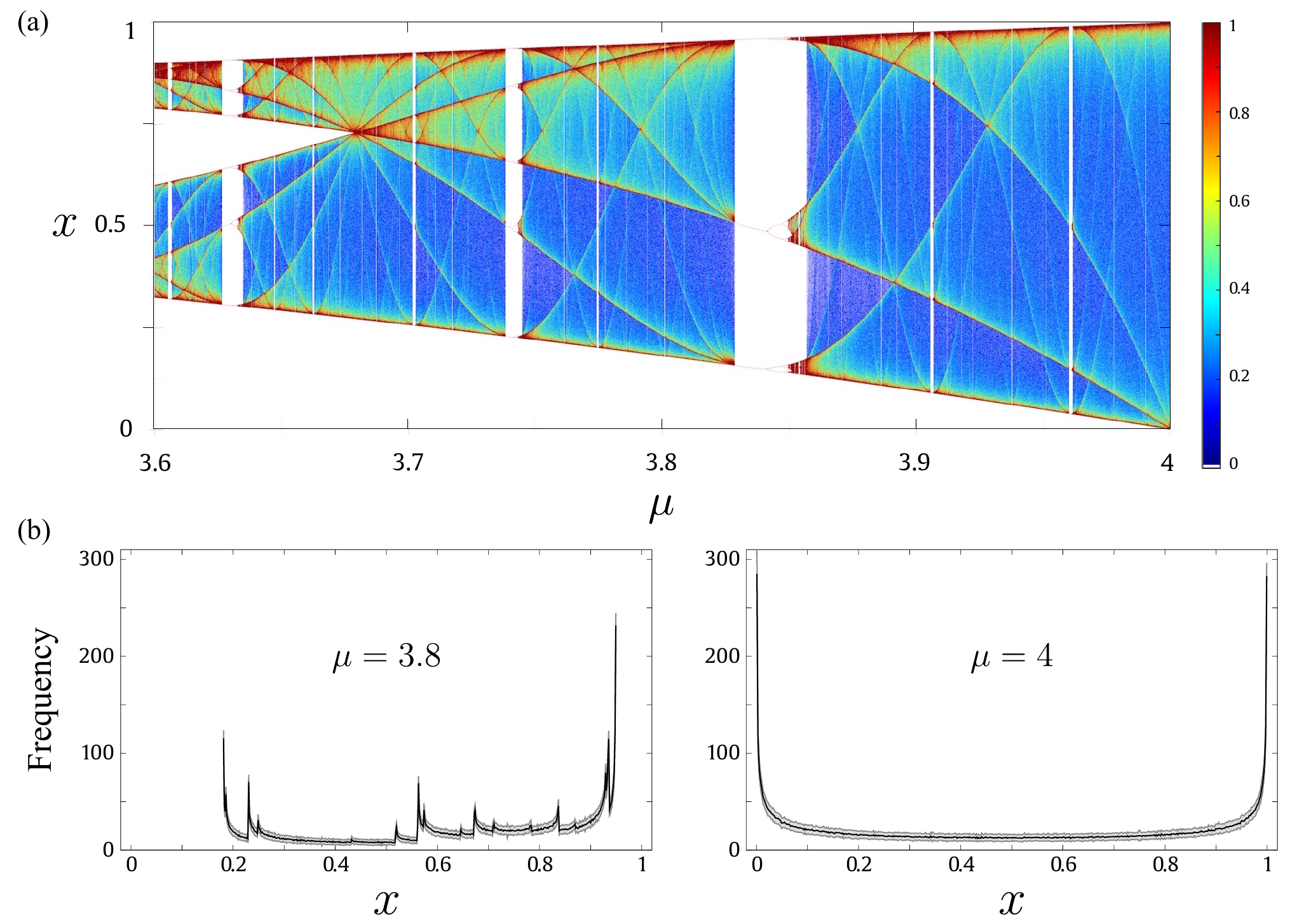} 
	\caption{a) Colored bifurcation diagram of the logistic map (Eq.~\eqref{eq:logisticmap}) depicted $\mu \in [3.6,4]$ in the x-axis, while the possible long-term values is shown in the y-axis, starting from the same initial condition, calculated over $10^5$ iterations (first $200$-th are the transient). b) Frequency distribution curve of the logistic map using $\mu=3.8$ and $\mu=4$, left and right, respectively. The x-axis shows the $x \in [0, 1]$ (500 bins) and the y-axis shows the frequency of the $10^4$ values discarding the first $10^3$ transient values. The curve represent the mean and standard deviation (shaded error bar) for sequences generated over 100 random initial conditions.}
	\label{fig:biffhistologmap} 
\end{figure*}

Even when the logistic map is considered as a source of randomness for some chaos-based PRNGs \cite{Chaos-PRNG9,Chaos-PRNG11,Chaos-PRNG12,Chaos-PRNG15}, its ergodicity introduce some weaknesses to the PRNGs that are based on it \cite{AlvarezCriptonalise, Persohn2012}. These limitations can be summarized in four main points: 

\begin{enumerate}[i]
\item Undesirable statistical features, e.g. non-uniform distribution;
\item Predictability, that is a consequence of a return time with memory;
\item Periodicity, which is inherent to chaos but that become an enhanced problem when dealing with finite numerical precision. Trajectories that should lead to long periodic cycles turn into short cycles \cite{Persohn2012}, consequence of the dynamical degradation of digital chaos \cite{DynamicalDegradation};
\item The quality of the pseudo-random sequence generated by a chaotic orbit. These weaknesses of chaos generated by the logistic map in the context of cryptography have been exposed in \cite{AlvarezCriptonalise,Persohn2012}. 
\end{enumerate}

There is an additional feature of chaos that is undesirable for cryptography, and that can also reflect on the characteristics of return times, so important for the Baptista's approach. The probability density of trajectories is typically not uniform. In Fig.~\ref{fig:biffhistologmap}(a), we show a bifurcation diagram for the parameter $\mu \in [3.6, 4]$. In this figure the colors represent the density with which points fall in a compartment. In Fig.~\ref{fig:biffhistologmap}(b) it is shown the frequency distribution of the logistic map using two parameters $\mu=3.8$ and $\mu = 4$ (corresponding to the rightmost region of the bifurcation diagram). Focusing on the parameter $\mu=4$, it can be observed that the frequency distribution has a classical ``U'' pattern. This frequency distribution diagram follows the invariant probability density function of the logistic map \cite{OttChaosBook} given by 

\begin{equation}
\label{eq:Ulogmap}
\rho(x) = \frac{1}{\pi\sqrt{x(1 - x)}}.
\end{equation}

In this particular case, a higher frequency can be observed at the edges $x \rightarrow 0$ and $x \rightarrow 1$ than in the central region. This map, if not properly encoded, could generate a PRNGs with non-uniform probabilities. In the case of the chaos ergodicity-based algorithm, the probability distribution of return times is not a plateau distribution as desired \cite{AlvarezCriptonalise}. Moreover, returns composing the cyphertext might have their average value inversely proportional to the invariant measure (given by $\textstyle\int_{x \in \epsilon} \rho(x)dx$) of the interval $\epsilon$ being used to constructed the return times (Kac's lemma) \cite{Baptista}. This would allow the identification of the control parameter being used to generate returns, leading to at least partial determination of the encoding sequence \cite{AlvarezCriptonalise}. 

The issues of the logistic map as source of randomness are related to the deterministic nature of the chaotic orbits and its ergodicity. In order for the logistic map to be appropriately used to generate random sequences, an additional analysis must be done. The weaknesses mentioned above are evident because the algorithms that use the numbers generated by the logistic map do not take into consideration all the infinitely long less significant digits of orbit points of the chaotic trajectories, these digits have infinitesimal depth \cite{machicao2017}. 

\subsection{k-logistic map: the deep-zoom approach}

The concept of deep-zoom into chaos was recently introduced by Machicao \& Bruno \cite{machicao2017}. It was shown that the dismissed digits after the decimal point can be advantageously used to improve the randomness quality of chaotic systems. This algorithm removes the first $k$-digits of the fractional part of each point of an orbit in order to compose a new orbit coming from an original orbit of a chaotic system. For instance, the $k$-logistic map is formed by considering the $k$-right digits of each $x^{t}$ value from an underlying orbit $k=0$ (generated originally by the logistic map). 

Given an orbit $\mathcal{O}(\mu, x_0)=\{x_0, x_1,\ldots, x_t\}$ as a series of points obtained from the logistic map, Eq.~\eqref{eq:logisticmap}, with parameters $\mu$ and $x_0$. The $k$-logistic map is defined as a function given by Eq.~\eqref{eq:klogistic} of the form $\phi_{k}(x_t):[0,1]\rightarrow ]0,1[$. Observe that the domain of the $k$-logistic map phase-space is $]0,1[$. This function generates orbits on the form $\mathcal{O}^{k}(\mu,x_0)= \{x_0^k, x_1^k,\ldots, x_t^k\}$, that is the resulting orbit derived from an underlying orbit $\mathcal{O}(\mu,x_0)$. The new values $x_t^k$ are formed by retaining $k$-th decimals to the right of the decimal separator corresponding to the underlying point $x_t \in \mathcal{O}^{k=0}$, a transformation that can be represented by 

\begin{equation}
x_t^k= \phi_{k}(x_t) = x_t 10^{k} - \lfloor x_t 10^k \rfloor\,,
\label{eq:klogistic}
\end{equation}
where $\lfloor$ $\rfloor$ stands for the floor function, and $x_0 \in ]0,1[$. Machicao \& Bruno \cite{machicao2017} recommended to use $\mu \rightarrow 4$, since the closer $\mu$ is to 4, the less dense are the periodic windows, and therefore, a parameter taken by chance has a larger positive probability of generating a chaotic orbit. To illustrate an output of Eq. (\ref{eq:klogistic}), observe two values generated by the logistic map: $x_{t}=0.581234$ and $x_{t+1}=0.523674185238$ corresponding to an orbit using $\mu=4$. To generate a $k$-logistic map orbit with $k=1$, indicates to discard the first position of the decimal places of $x_{t}$, i.e. $\phi_{k=1}(x_{t})= 0.81234$, and analogously $\phi_{k=3}(x_{t+1})= 0.6744185238$.

\begin{figure}[!hb]
	\centering
	\begin{center}
		\includegraphics[scale=0.84]{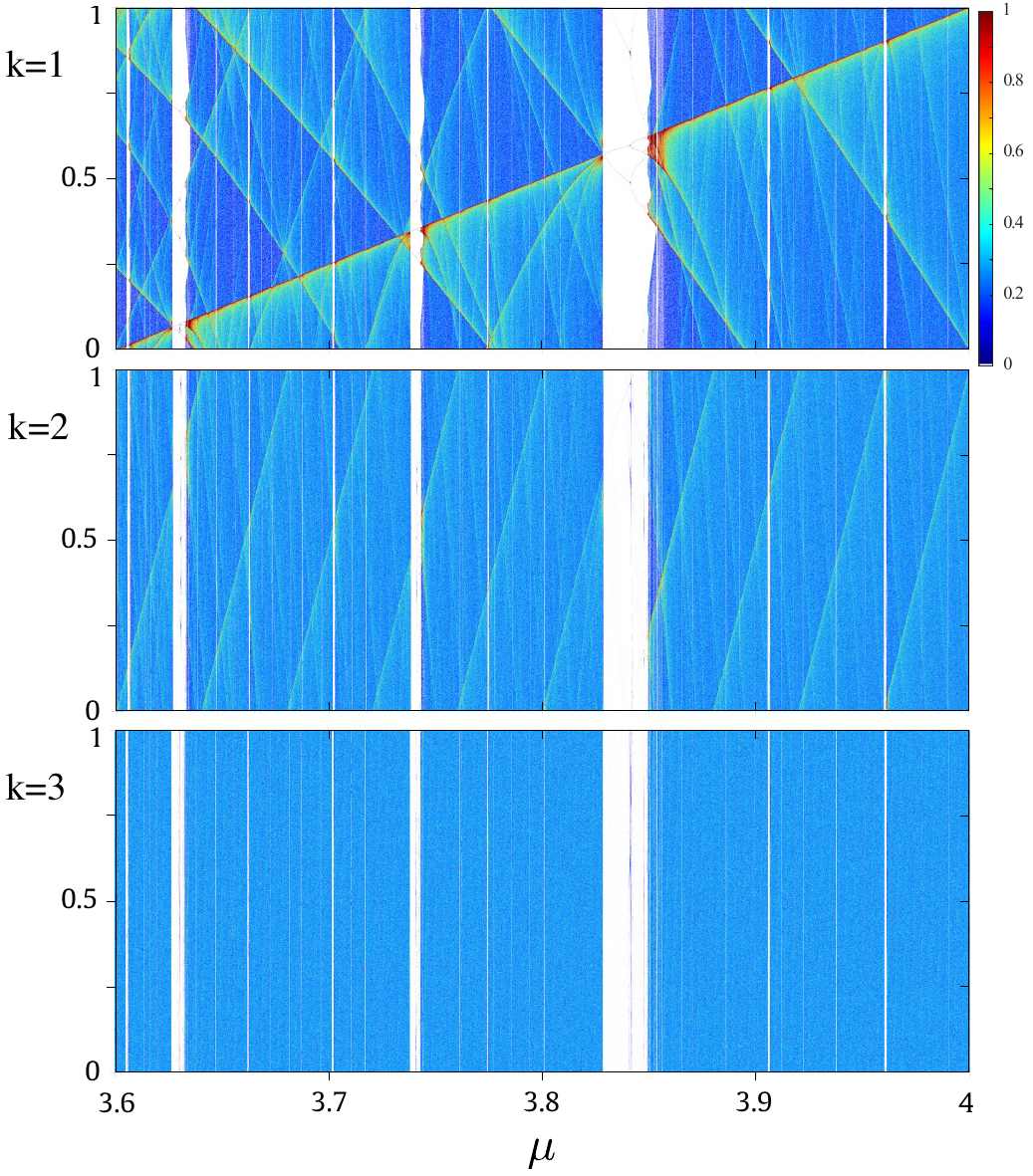}
		\caption{Colored bifurcation diagrams of the $k$-logistic map (Eq.~\eqref{eq:klogistic}) for $k=1$ (upper panel), $k=2$ (middle panel) and $k=3$ (lower panel), from top to down. The x-axis depicts $\mu \in [3.6,4]$, while the y-axis shows the $k$-logistic map, starting from the same initial condition, calculated over $10^5$ iterations (first $200$-th are the transient).}
		\label{fig:biffklogmap}
	\end{center}
\end{figure}

From Fig.~\ref{fig:biffklogmap} an interesting phenomenon can be observed when comparing the original bifurcation diagram of the logistic map shown in Fig.~\ref{fig:biffhistologmap}(a) with the bifurcation diagrams of the $k$-logistic map with $k=1$, $k=2$, and $k=3$. It is observed that the chaotic patterns interleaved predominantly in the $\mu \in [3.57, 3.828]$ and $\mu \in [3.86, 4]$ regions and that follow a zigzag pattern ($k = 1$) begin to disappear gradually as $k$ is enlarged. No apparent pattern for the trajectory's probability distribution is observed for $k=3$. In fact, the distribution frequency of the $k$-logistic map becomes rapidly uniform, even for moderately large $k$ (see Figs.~\ref{fig:histogramas}(b-d)).

\section{Proposed chaos-based cryptographic system}\label{sec4}

The deep-zoom approach has shown interesting results. In fact, it was shown that the randomness quality is improved as $k$ is increased, because of its rapid transition to ``strong'' randomness as $k$ tends to infinity, which was demonstrated throughout a set of empirical tests \citep{machicao2017}. Therefore, what would happen when the logistic map of the Baptista's algorithm is replaced by the $k$-logistic map?

Recall the block diagrams shown in Fig.~\ref{fig:baptistaAlgorithm} where the function $f(x_n)$ inside the encryption/decryption blocks corresponds to Eq.~\eqref{eq:logisticmap}. Our proposed cryptosystem simply embed the $k$-logistic map function $\phi_k(x_n)$ given by Eq.~\eqref{eq:klogistic} into the original approach. In fact, one could embed almost any chaotic function to generate returns. For the purposes of comparison, we will also consider in Sec.~\ref{sec:analysis} a general PRNG to extract first returns. Notice however that chaotic functions are usually advantageous in contrast to general PRNGs, since they usually require less computational power and are quicker to be implemented. 

In order to exemplify the encryption process using the $k$-logistic map, we consider the message ``hi'', as shown in our previous example of Section~\ref{sec:Baptista}, and same parameters $\mu=3.8$, $N_0=250$ transient, $x_0=0.232323$, $\eta=0$. These ciphertexts are reported in Table~\ref{tab:encryptingklogmap} for each parameter $k=0$ (original logistic map), $k=1$, $k=2$, $k=3$ and $k=4$.

\begin{table}[!ht]
	\centering
	\caption{Encryption procedure of the ergodicity chaos-based algorithm using the $k$-logistic map from top to down $k=0$, $k=1$, $k=2$ and $k=3$. From left to right the plaintext ``hi'' associated to their corresponding site number based on an initial condition $x_0$, which is iterated over $x$ time steps until returns a ciphertext $C_n$.}
	\label{tab:encryptingklogmap}
	\tiny
		\begin{tabular}{llllll}
			\toprule			
			$k$&Plain & $x_0$ & $x_n$ & Site & $C_n$\\
			\hline
			\addlinespace[1ex] 
			\multirow{2}{*}{$k=0$}&``h'' & 0.23232300000000 & 0.44160905447136& 104 & 1713 \\
			&``i'' & 0.44160905447136 & 0.44486572362642& 105 & 364 \\
			\hline
			\addlinespace[1ex] 
			\multirow{2}{*}{$k=1$}&``h'' & 0.23232300000000 & 0.44194259087588& 104 & 529 \\
			&``i'' & 0.44194259087588 & 0.44472392303434& 105 & 573 \\
			\hline
			\addlinespace[1ex] 
			\multirow{2}{*}{$k=2$}&``h'' & 0.23232300000000 & 0.44306123881144& 104 & 609 \\
			&``i'' & 0.44306123881144 & 0.44499232862576& 105 & 1671 \\
			\hline
			\addlinespace[1ex] 
			\multirow{2}{*}{$k=3$}&``h'' & 0.23232300000000 & 0.44192450305195& 104 & 1361 \\
			&``i'' & 0.44192450305195 & 0.44586158032371& 105 & 892 \\
			\hline
			\addlinespace[1ex] 
			\multirow{2}{*}{$k=4$}&``h'' & 0.23232300000000 & 0.44174262765409& 104 & 592 \\
			&``i'' & 0.44174262765409 & 0.44401200078573& 105 & 399 \\
	\bottomrule		
		\end{tabular} 
\end{table}


\section{Analyzing the cryptographic quality}
\label{sec:analysis} 
We now analyze the quality of the proposed method based on identified weaknesses of the \citeauthor{Baptista} algorithm. This section is dedicated to analyse the good properties of the $k$-logistic map, with respect to its frequency distribution, the frequency distribution of its ciphertext, the strong non-invertible property of its return-map, and its pseudo-randomness properties.

\subsection{Uniform probability distribution function}

\begin{figure}[!ht]
	\centering
	{\includegraphics[scale=0.82]{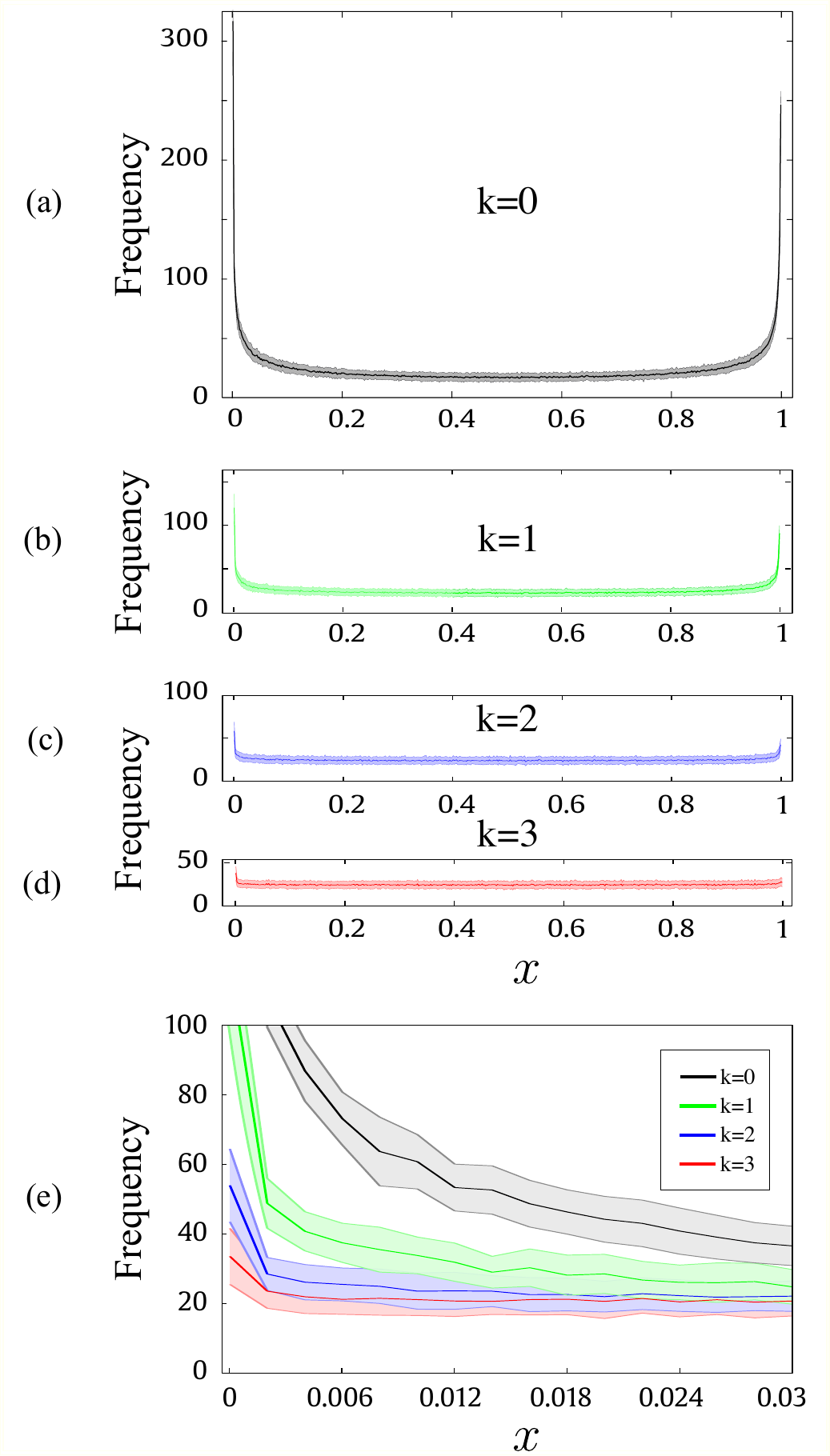}}
	\caption{Frequency distribution curves for (a) the original logistic map, (b) $k=1$, (c) $k=2$ and (d) $k=3$ using $\mu=4$. The horizontal axis shows the $x \in [0, 1]$ (500 bins) and the vertical axis shows the frequency of the $10^4$ values discarding the first $10^3$ transient values. The curves represent the mean and standard deviation (shaded error bar) for sequences generated over 100 random initial conditions. e) The inset plot depicts a zoom on the windows $x \in [0, 0.03]$ for these 4 plots.} 
	\label{fig:histogramas}
\end{figure}

In order to avoid an attacker to make predictions about the plaintext, the distribution of the points of an orbit of a chaotic function is expected to be uniform. In the original chaos based algorithm, the relevant distribution is the one from the return times, which however depends on the properties of the distribution of the chaotic trajectory. As can be seen in Fig.~\ref{fig:histogramas}(a), the probability distribution function of the logistic map is not uniform. However, as the logistic map is analyzed into the deep-zoom, the distribution tends to become increasingly uniform, as can be observed in Figs.~\ref{fig:histogramas}(b-d), for $k =1$ (green), $k=2$ (blue) and $k =3$ (red), respectively. Figure ~\ref{fig:histogramas}(e) shows these densities for a region as $x$ approaches to zero. We see that the $k$-logistic map resolves security issues that would emerge as a consequence of the non-uniformity of the distribution.

Thus, the $k$-logistic map resolves the \textbf{weakness (i)}, about undesirable statistics of the chaotic sequence.

\subsection{Ciphertext distribution}

\begin{figure*}[!ht]
	\centering
	{\includegraphics[scale=0.8]{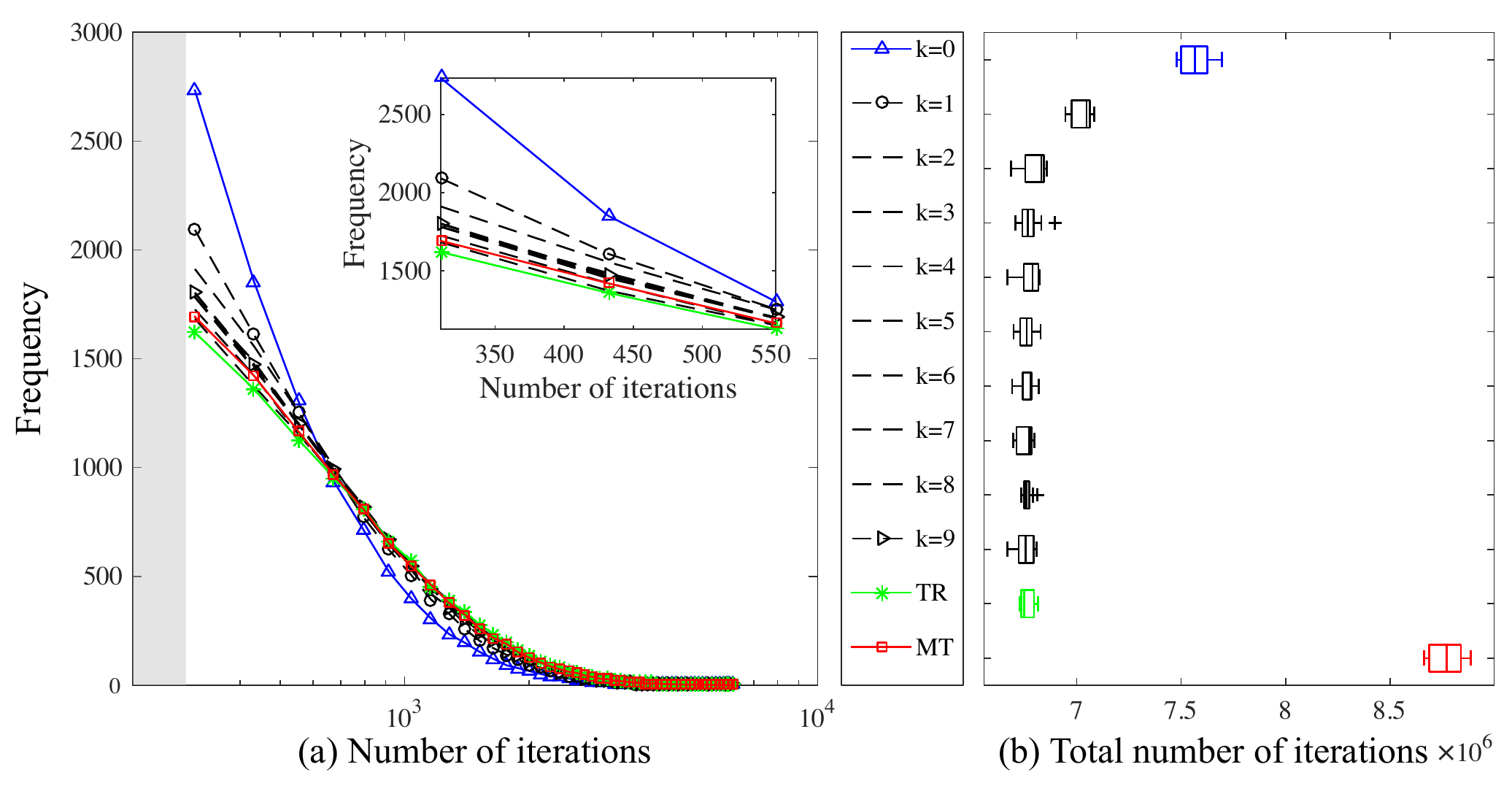}}
	\caption{(a) Average ciphertext distribution for the original method ($k=0$), the $k$-logistic map with different parameters $k=\{1,2,\ldots,9\}$, the Mersenne Twister PRNG (identified by ``MT'' in figure legend) and true randomness sources from atmospheric noise (identified by ``TR'' in figure legend). The x-axis shows a logarithmic scale of the number of iterations (ciphertext) used to encrypt each of the $10^4$ random letters. The average curve was calculated over 100 randomly chosen plaintexts. The gray region correspond to the transient time $N_0 = 250$ iterations. The inset figure depicts a small region from the figure. (b) Boxplot of the total number of iterations corresponding to the 100 plaintexts samples.}
	\label{fig:histCiphertext}
\end{figure*}

Here, we analyze the frequency distribution of the return times (ciphertext) used to encrypt a plaintext. We performed an experiment to compare the distribution from the original method and the proposed improved method with the $k$-logistic map $k=\{1,2,\ldots,9\}$. Moreover, we also considered to explore the characteristics of the ciphertexts using a commercial PRNG, the Mersenne Twister PRNG \cite{MersenneTwister,MersenneJava} and a true random source \cite{RANDOM-ORG} sequences of numbers converted from the atmospheric noise that were download from \cite{RANDOM-ORG}, as the generators of the return times in the encryption/decryption method, instead of using the logistic map. 

In this experiment we encrypted 100 randomly chosen plaintexts containing $10^4$ letters using the proposed method with the $k$-logistic map $k=\{1,2,\ldots,9\}$ and with the original logistic map. The average frequency distribution of the ciphertexts is shown in Fig.~\ref{fig:histCiphertext}(a). The average is calculated considering 100 different plaintexts. From this figure, we can observed that the curves for the proposed method tend to get close to the true randomness curve as $k$ increases. Moreover, in Fig.~\ref{fig:histCiphertext}(b), it is shown a boxplot figure corresponding to the accumulative number of iterations corresponding to the 100 ciphertexts. From this figure, we can observe that the bigger $k$ is, the shorter the time-series required to encrypt. Thus, the larger $k$ is, the less computational cost required. Moreover, our method has a better distribution as compared to that of the commercial counterpart Mersenne Twister.

Thus, the $k$-logistic map resolves the \textbf{weakness (i)}, about undesirable statistics of the cyphertext.
\subsection{Shuffling return map}

\begin{figure}[!ht]
	\centering
	{\includegraphics[scale=0.82]{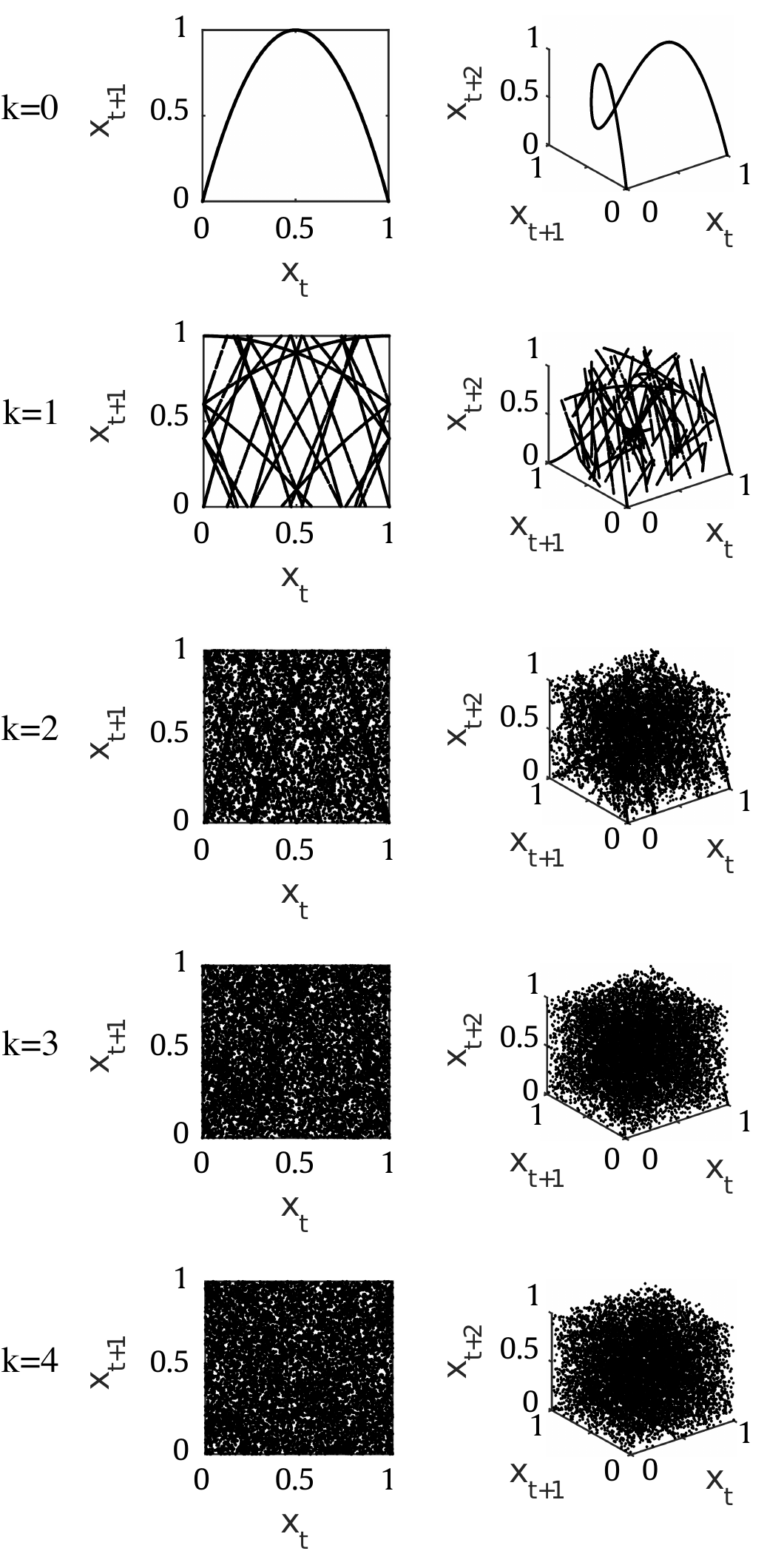}}
	\caption{Phase diagrams for the $k=0$, $k=1$, $k=2$, $k=3$ and $k=4$-logistic map using $\mu = 4$. Two- and three-dimensional diagrams are shown in the left and right columns, respectively. The horizontal and vertical axes show the phase space of $x^k_t$ against $x^k_{t+1}$. Each orbit contains $10^4$ points starting from random initial conditions, where the first 200 iterations were discarded (transient time).}
	\label{fig:poincareLogistic}
\end{figure}

When the return map of the logistic map is plotted it is noticed the well-known inverted parabola function, as can be seen in the upper left panel in Fig.~\ref{fig:poincareLogistic}, for $k=0$. Its 3D visualization in the upper right panel also revels its standard continuous pattern, describing a strong correlation between past and future iterates. Maps that generate distinct pattern on the return map can lead to information about the ciphertext. It is to be expected that a return map of uncorrelated sequences (e.g. as those generated by PRNGs) would densely populate the space, in a demonstration that the lost of correlation. The $k$-logistic map destroys these typical patterns. 

The classical inverted parabola disappears when $k=1$ (second row in Fig.~\ref{fig:poincareLogistic}), producing a set of discontinuous curves, producing a large number of period-1 orbits. As $k$ increases, the points are scattered in the phase diagram and any visual pattern seems to disappear as can be seen in the diagrams for $k=2$ (third row panels) to $k=4$ (lower row panels). As $k$ increases, the phase diagram tends to look like more of one produced by uncorrelated sources, with a map that is highly discontinuous and also highly non-invertible. 

So, the use of the $k$-logistic map should ensure an increase in the security level of the proposed ergodic cryptosystem, since it produces trajectories that can be hardly predictable, this contributing to resolve the \textbf{weakness (ii)} of the logistic map to the ergodic cryptosystem. It also handles well the problem of long periodic orbits (\textbf{weakness iii}) being turned into lower cycles, by actually creating an infinitude of lower periodic cycles.

\subsection{Succeed pseudo-randomness tests}

A final weakness of the ergodic cryptosystem regards the pseudo-random quality of the chaotic trajectory (\textbf{weakness (iv)}). Here, we cryptanalyse the chaotic sequences generated by the $k$-logistic map to show that this weakness would be resolved by the use of the k-logistic map in the ergodic cypher. 

Despite the extensive use of PRNGs in business and in scientific research, there is a lack of manners to analyze the randomness quality of the cryptographic methods. What is commonly used today to measure and test the pseudo-random sequences are the statistical test suites. The scientific literature highlights two of them: NIST and DIEHARD. The objectives of these tests are to analyze whether the sequences generated by PRNGs present patterns or any regular behavior. These statistical tests are limited in gathering evidence that the PRNG generated numbers are indeed from a truly random source. However, they are a first estimation, which we will adopt in this work.  

\newcommand{\pp}{ \cellcolor[HTML]{C0C0C0}} %
\begin{table*}[!ht]
	\caption{Average number of files that passed DIEHARD tests using the $k$-logistic map PRNG (Eq.~\eqref{eq:klogistic}) from 100 file samples. Severely failed tests are shown in gray. All tests passed using the interval $0.0001<\text{p-value}<0.9999$. Reproduced from \cite{machicao2017}.}
	\label{tab:diehard}
	\begin{tabular}{ll}
	\end{tabular}
	\fontsize{5.5}{5.6}\selectfont
	\centering 
	\begin{tabular}{lrrrrrrrrrr}
		\toprule			
		\textbf{DIEHARD tests}  & \scalebox{0.65}[1]{$k=0$} & \scalebox{0.65}[1]{$k=1$}& \scalebox{0.65}[1]{$k=2$}& \scalebox{0.65}[1]{$k=3$}& \scalebox{0.65}[1]{$k=4$}& \scalebox{0.65}[1]{$k=5$}& \scalebox{0.65}[1]{$k=6$}& \scalebox{0.65}[1]{$k=7$}& \scalebox{0.65}[1]{$k=8$}& \scalebox{0.65}[1]{$k=9$}\Tstrut\\
		\hline \Tstrut
		BirthdaySpacings [KS] & 100 & 100 & 100 & 100 & 100 & 100 & 100 & 100 & 100 & 100 \\
		OverlappingPermutations & 99 & 97 & 98 & 95 & 98 & 96 & 98 & 98 & 99 & 100 \\
		Ranks31x31 matrices & 100 & 100 & 100 & 100 & 100 & 100 & 100 & 100 & 100 & 100 \\
		Ranks32x32 matrices & 100 & 100 & 100 & 100 & 100 & 100 & 100 & 100 & 100 & 100 \\
		Ranks6x8 matrices [KS] & \pp0 & \pp0 & \pp25 & 99 & 100 & 100 & 100 & 100 & 100 & 100 \\
		Monkey20bitsWords [KS] & \pp0 & 99 & 100 & 100 & 100 & 100 & 100 & 100 & 100 & 100 \\
		OPSO [KS] & 98 & 99 & 100 & 100 & 100 & 100 & 100 & 100 & 100 & 100 \\
		OQSO [KS] & 98 & 100 & 100 & 100 & 100 & 100 & 100 & 100 & 100 & 100 \\
		DNA [KS] & 100 & 100 & 100 & 100 & 100 & 100 & 100 & 100 & 100 & 100 \\
		Count1sStream & \pp0 & \pp0 & \pp0 & 98 & 100 & 100 & 100 & 100 & 100 & 100 \\
		Count1sSpecific [KS] & \pp0 & \pp0 & \pp0 & \pp0 & 94 & 100 & 100 & 100 & 100 & 100 \\
		ParkingLot [KS] & 100 & 100 & 100 & 100 & 100 & 100 & 100 & 100 & 100 & 100 \\
		MinimumDistance [KS] & 96 & 100 & 100 & 100 & 100 & 100 & 100 & 100 & 99 & 100 \\
		RandomSpheres [KS] & 100 & 100 & 100 & 100 & 100 & 100 & 100 & 100 & 100 & 100 \\
		Squeeze [KS] & 100 & 100 & 100 & 100 & 100 & 100 & 100 & 100 & 100 & 100 \\
		OverlappingSums [KS] & 100 & 100 & 100 & 100 & 100 & 100 & 100 & 100 & 100 & 100 \\
		Runs (up) & 100 & 100 & 100 & 100 & 100 & 100 & 100 & 100 & 100 & 100 \\
		Runs (down) & 100 & 100 & 100 & 100 & 100 & 100 & 100 & 100 & 100 & 100 \\
		Craps (wins) & 100 & 100 & 100 & 100 & 100 & 100 & 100 & 100 & 100 & 100 \\
		Craps (throws/game) & 100 & 100 & 100 & 100 & 100 & 100 & 100 & 100 & 100 & 100 \\
		\bottomrule
	\end{tabular}
\end{table*}

In order to ascertain that, Table \ref{tab:diehard} and Table \ref{tab:nist} are reproduced from the original work \cite{machicao2017}. Generally, empirical tests are addressed for PRNGs in order to analyze the mathematical properties of the sequences. The tests were performed in the DIEHARD \cite{DIEHARD} and NIST \cite{NIST-PRNG} test suites. These tests formulate hypotheses to verify if the distribution of the random sequence of entry is adhered to some known distribution.

Here we propose to evaluate the quality of the random source of the ergodic cryptographic system, which boils down to evaluating the $k$-logistic map for several values of the $k$ parameter. Thus, 100 files were generated, each containing $28 \times 10^5$ (12.5 MB) numbers generated for each $k=\{0,1, \ldots, 9 \}$. Note that the file size used is due to the constraints of DIEHARD \cite{DIEHARD} itself. Each sample was generated from a random seed, and each value $x_{t}^k$ of each orbit was transformed into integers using a $32$-bit mask (discretization).

The results of the 18 statistical tests are reported in Table \ref{tab:diehard} for the DIEHARD suit and in Table \ref{tab:nist} for the NIST suit. Each column corresponds to the number of files that passed the sub-tests. In both tables, the tests that failed in at least 50 files were highlighted in gray, which can be observed in the case of $k=0$, $ k=1$, $ k=2$ and $k=3$-logistic map. 
As expected, $k=0$ fails to both DIEHARD and NIST. The panorama changes as the parameter $k$ increases. The $k$-logistic map passes on all of the tests from DIEHARD and NIST when $k\geq 4$. Accordingly, the ergodic cypher should not present weakness with respect to the pseudo-random nature of the source, in this case the $k$-logistic map.

\setlength{\tabcolsep}{2pt}
\begin{table}[!ht]
	\caption{Number of files that passed the NIST test suites \cite{NIST-PRNG} for the $k$-logistic map. Failed tests are shown in gray. All the tests passed to the $\alpha=0.01$ significance level. Reproduced from \cite{machicao2017}.}
	\label{tab:nist}
	\fontsize{5.5}{5.6}\selectfont
	\centering
		\begin{tabular}{lrrrrrrrrrr}
			\toprule			
			NIST test & \scalebox{0.65}[1]{$k=0$} & \scalebox{0.65}[1]{$k=1$}& \scalebox{0.65}[1]{$k=2$}& \scalebox{0.65}[1]{$k=3$}& \scalebox{0.65}[1]{$k=4$}& \scalebox{0.65}[1]{$k=5$}& \scalebox{0.65}[1]{$k=6$}& \scalebox{0.65}[1]{$k=7$}& \scalebox{0.65}[1]{$k=8$}& \scalebox{0.65}[1]{$k=9$}\\
			\hline
			\Tstrut
			Frequency & 98 & 99 & 99 & 99 & 99 & 99 & 99 & 99 & 99 & 99 \\
			\Tstrut
			BlockFrequency $(m=128)$ & \pp{0} & \pp{1} & 66 & 95 & 98 & 98 & 99 & 100 & 99 & 99 \Tstrut\\
			\hline 
			\multicolumn{11}{l}{CumulativeSums} \Tstrut\\
			\hspace{1cm}Forward sums & 97 & 98 & 99 & 99 & 98 & 99 & 99 & 99 & 99 & 99 \\
			\hspace{1cm}Reverse sums & 97 & 99 & 99 & 99 & 99 & 99 & 99 & 99 & 99 & 99 \\
			\hline 
			\Tstrut
			Runs &\pp{0} & \pp{0} & \pp{14} & 91 & 98 & 99 & 99 & 99 & 99 & 100 \\
			\Tstrut
			LongestRun & \pp{0} & \pp{0} & \pp{15} & 89 & 98 & 99 & 98 & 100 & 99 & 99\\
			\hline
			\Tstrut
			Rank & 99 & 100 & 99 & 99 & 99 & 99 & 99 & 99 & 99 & 99 \\
			\hline
			\Tstrut
			FFT & 77 & 98 & 99 & 99 & 99 & 99 & 99 & 99 & 99 & 99\\
			\hline 
			\multicolumn{11}{l}{Non-overlappingTemplate} \Tstrut\\
			\hspace{1cm}000000001 & \pp{0} & \pp{0} & \pp{48} & 97 & 99 & 99 & 99 & 100 & 99 & 99 \\
			\hspace{1cm}000000011 & \pp{0} & \pp{3} & 87 & 98 & 99 & 99 & 99 & 99 & 99 & 99 \\
			\hspace{1cm}000000101 & \pp{0} & \pp{41} & 94 & 98 & 98 & 99 & 99 & 99 & 98 & 99 \\ 
			\hline
			\Tstrut
			OverlappingTemplate & \pp{0} & \pp{0} & \pp{11} & 93 & 98 & 99 & 98 & 99 & 99 & 99 \\
			\hline
			\Tstrut
			Universal & \pp{0} & 68 & 97 & 98 & 99 & 99 & 99 & 99 & 99 & 99 \\
			ApproxEntropy $(m=10)$ & \pp{0} & \pp{0} & 64 & 98 & 99 & 99 & 99 & 100 & 99 & 99 \\
			\hline
			\multicolumn{11}{l}{RandomExcursions}\Tstrut \\
			\hspace{1cm}$x= $ -4 & 90 & 98 & 99 & 99 & 98 & 99 & 99 & 99 & 99 & 100 \\
			\hspace{1cm}$x=$ -3 & 91 & 97 & 99 & 99 & 99 & 99 & 99 & 99 & 99 & 99 \\
			\hspace{1cm}$x=$ -2 & 94 & 99 & 98 & 99 & 99 & 98 & 99 & 99 & 99 & 99 \\
			\hspace{1cm}$x=$ -1 & 95 & 99 & 99 & 98 & 99 & 99 & 99 & 99 & 99 & 100 \\ 
			
			\hline
			\multicolumn{11}{l}{RandExcursVar} \Tstrut\\
			\hspace{1cm}$x=$ -9 & 99 & 100 & 99 & 99 & 100 & 99 & 99 & 99 & 99 & 100 \\
			\hspace{1cm}$x=$ -8 & 99 & 99 & 99 & 99 & 99 & 99 & 99 & 99 & 99 & 100 \\
			\hspace{1cm}$x=$ -7 & 100 & 99 & 99 & 99 & 99 & 99 & 99 & 99 & 99 & 100 \\
			\hspace{1cm}$x=$ -6 & 100 & 99 & 99 & 99 & 99 & 99 & 98 & 99 & 99 & 99 \\
			\hspace{1cm}$x=$ -5 & 100 & 99 & 99 & 99 & 99 & 99 & 99 & 99 & 99 & 99 \\
			\hspace{1cm}$x=$ -4 & 99 & 100 & 99 & 99 & 99 & 99 & 99 & 99 & 99 & 99 \\
			\hspace{1cm}$x=$ -3 & 99 & 100 & 99 & 98 & 99 & 99 & 99 & 99 & 99 & 99 \\
			\hspace{1cm}$x=$ -2 & 99 & 99 & 99 & 98 & 99 & 99 & 99 & 100 & 99 & 99 \\
			\hspace{1cm}$x=$ -1 & 99 & 99 & 99 & 99 & 99 & 99 & 99 & 99 & 99 & 99 \\ 
			\hline
			\multicolumn{11}{l}{Serial $(m=16)$} \Tstrut\\
			\hspace{1cm} Serial 1 & \pp{0} & \pp{1} & 82 & 96 & 98 & 99 & 99 & 98 & 99 & 99 \\
			\hspace{1cm} Serial 2 & \pp{10} & 81 & 95 & 98 & 98 & 99 & 99 & 99 & 98 & 99\\
			\hline \Tstrut
			LinearComplexity $(M=500)$ & 99 & 98 & 99 & 99 & 99 & 99 & 99 & 98 & 99 & 99 \\
\bottomrule			
		\end{tabular} 
\end{table}

\section{Discussions and conclusion}\label{sec6}


This paper analyzed a classic algorithm published in 1998 in a more sophisticated way. The original algorithm has been much criticized in the literature, due to both its computational cost and its theoretical weaknesses. In this paper, we study a cryptosystem based on \citeauthor{Baptista}'s original ergodic cryptosystem algorithm, but that uses a chaotic trajectory that is created by the deep-zoom $k$-logistic map. This paper is not intended at constructing a cryptographically safe algorithm, but instead to comprehensively study how the security of the ergodic cryptosystem would be enhanced if embedded with the $k$-logistic map. Some of our analysis were carried out in the generated cyphertext, such as the statistics of the return times. Others were carried out in the chaotic source, the $k$-logistic map. \citeauthor{Baptista}'s original algorithm has the number of iterations as the ciphertext, which can not be directly used for randomness tests based on the DIEHARD and NIST test suits. These tests were therefore performed in the chaotic generator used. More than trying to create a cryptosystem that can be commercially exploited, we sought not only to improving the original approach but also to open a new branch of study for the chaos-based cryptosystem community.  

Some considerations need to be taken into account. The first point concerns computational cost. In general, the cost is high for both the ergodic cryptosystem and our proposed method. However, we must salient that, once the orbit is calculated, the $k$-th values of the generated orbits do not increase the computational cost as $k$ gets larger, since the performance relies mostly on the first calculation of the original orbit. Importantly, as $k$ gets larger the security improves. Another point is that the original ergodic cryptosystem uses double precision, which indeed, due to round-off errors it may quickly fall into a periodic window induced by round-off errors. The FORTRAN double precision applied in the original chaos based algorithm guarantees 15 digits of accuracy and a magnitude ranging of 10 from -308 to +308. We fixed this issue by using high arbitrary precision arithmetic. A high precision can be achieved using the high-performance arbitrary precision arithmetic library Apfloat\footnote{http://www.apfloat.org/}. It performs calculations with a precision of millions of digits, providing true chaotic, and therefore aperiodic orbits. 

In the original ergodic cryptosystem, the round-off operations could lead to a non-invertible encryption procedure. This problem is evident in the scenario where the computer of the message transmitted has greater precision than the computer of the receiver. The $k$-logistic map is implemented over a arbitrary precision arithmetic, then the round-off error can be used in our favor. For instance, the number of precision can be setup. Obviously, we should maintain a reasonable precision in order to guarantee the $k$-logistic map to produce randomness.

Regarding the conjugacy, the logistic map at $\mu=4$ is topologically conjugated to the tent map as well as the shift map \cite{stojanovski2001chaos}. Due to the fact that these maps have the same dynamical behavior, there is some possibility to do inverse engineering. The $k$-logistic map may also present this issue. So, we recommended using $\mu \rightarrow 4$, in order to avoid conjugacy with tent map or shift map, but still being able to create a trajectory with higher entropy. 

We hope this work will create a challenge for the scientific community, 
opening new avenues in which research will be conducted to search for possible deficiencies or further developments in the proposed improved cryptosystem. For example, attempts to discover patterns in the proposed PRNG sequences may lead to the appearance of new weaknesses in the proposed cryptosystem, revealing new possibilities for the field of cryptography. Scrutiny and new improvements might create after all a secure cryptographic systems out of the one studied in this work.

\section*{Acknowledgments}

J. M. gratefully acknowledges the financial support from the National Council for Scientific and Technological Development, Brazil (CNPq) grant \#405503/2017-2.
M. A. is grateful for the support of the Coordination for the Improvement of Higher Education Personnel (CAPES PROEX-9524331/M).
O. M. B. acknowledges support from CNPq (grant \#307797/2014-7) and FAPESP (grant \#14/08026-1).


\end{document}